\documentclass[aps,preprint,showpacs]{revtex4}
\usepackage{epsfig}
\usepackage{amsfonts}
\usepackage{amsmath}
\usepackage{amssymb}
\usepackage{graphicx}
\begin{document}
\preprint{LEZ/00103}
\title{Boundary Solutions of the Two-electron Schr\"{o}dinger Equation at
Two-particle Coalescences of the Atomic Systems}

\author{E.~Z.\ Liverts}
\affiliation{Racah Institute of Physics, The Hebrew University,
Jerusalem 91904, Israel}
\author{M.~Ya.\ Amusia}
\affiliation{Racah Institute of Physics, The Hebrew University,
Jerusalem 91904, Israel; A.~F.\ Ioffe Physical-Technical Institute,
St. Petersburg, 194021, Russia}
\author{R.\ Krivec}
\affiliation{Department of Theoretical Physics, J.\ Stefan Institute,
P.O. Box 3000, 1001 Ljubljana, Slovenia}
\author{V.~B.\ Mandelzweig}
\affiliation{Racah Institute of Physics, The Hebrew University,
Jerusalem 91904, Israel}

%\keywords{one two three}

\begin{abstract}
The limit relations for the partial derivatives of the two-electron
atomic wave functions at the two-particle coalescence lines have
been obtained numerically using accurate CFHHM wave functions. The
asymptotic solutions of the proper two-electron Schr\"{o}dinger
equation have been derived for both electron-nucleus and
electron-electron coalescence. It is shown that the solutions for
the electron-nucleus coalescence correspond to the ground and singly
excited bound states, including triplet ones. The proper solutions
at small distances $R$ from the triple coalescence point were
presented as the second order expansion on $R$ and $\ln R$. The
vanishing of the Fock's logarithmic terms at the electron-nucleus
coalescence line was revealed in the frame of this expansion, unlike
the case of electron-electron coalescence. On the basis of the
obtained boundary solutions the approximate wave function
corresponding to both coalescence lines have been proposed in the
two-exponential form with no variational parameters.
\end{abstract}

%%%%%%%%%%%%%%%%%%%%%%%%%%%%%%%%%%%%%%%%%%%%%%%%%%%%%%%%%%%%%%%%%%%%%%%%%%%%%%%

\maketitle
%------------------------------------------------
\section{Introduction}

Two-electron systems present an excellent basis both for testing the
new quantum calculational approaches to many-body systems and for
studying a number of photo-electron and other atomic processes. This
is because such systems are the simplest ones with enough complexity
to contain the main features of a many-body theory. This complexity
arises from the electron-electron Coulomb potential which depends on
the interelectronic distance \( r_{12}=\mid
\mathbf{r}_{1}-\mathbf{r}_{2}\mid  \). The proper Hamiltonian for
infinite nuclear mass and charge \( Z \), in atomic units used
throughout the paper, is given by
\begin{equation} \label{1}
H=-\frac{1}{2}(\nabla^{2}_{1}+\nabla^{2}_{2})-\frac{Z}{r_{1}}-\frac{Z}{r_{2}}+\frac{1}{r_{12}}.
\end{equation}
 It does not depend on any experimental constants whose values
change considerably with improvements in measurement equipment.
Therefore, it provides a standard for theoretical calibration.

It has been established \cite{1} that relativistic and
quantum-electrodynamic corrections to the energy levels of an atomic
or molecular system require highly accurate nonrelativistic wave
functions. Rayleigh-Ritz variational calculations provide a wave
function with relative error approximately proportional to the
square root of the relative error in the energy. Therefore, if the
energies are used to estimate the quality of the wave functions,
then it is necessary to calculate the nonrelativistic energies to
far greater accuracy than would otherwise be needed.

The alternative way for obtaining the very accurate and locally correct wave
functions is a direct solution of the three-body Schr\"{o}dinger equation. The Correlation
Function Hyperspherical Harmonic Method (CFHHM), employed in this paper, realizes
successfully this way of solution \cite{2}. Accuracy of the method is comparable
to the most sophisticated variational calculations.

For problems in atomic or molecular physics, eigenfunctions of the Hamiltonian (1)
exhibit singular behaviour at those points in configuration space where two
or more charged particles come together and the resulting potential becomes
infinite. For systems with zero total orbital momentum (\( S \)-states) the
wave function depends only on three scalar variables \( r_{1}, \) \( r_{2} \)
and \( r_{12} \), i. e., \( \Psi \equiv \Psi (r_{1},r_{2},r_{12}) \). At the
two-particle coalescences, the derivatives of the wave function \( \Psi  \)
have discontinuities characterized by the famous Kato cusp conditions \cite{3},
which have the simplest form for the \( S \)-state of a two-electron atomic system :
\begin{equation} \label{2}
\frac{\partial \Psi }{\partial r_{1}} \mid_{r_{1}=0} =-Z\Psi(0,R,R),
\ \ [r_{2}=r_{12}=R]
\end{equation}
\begin{equation} \label{3}
\frac{\partial \Psi }{\partial r_{2}} \mid_{r_{2}=0}
=-Z\Psi (R,0,R),\ \ [r_{1}=r_{12}=R]
\end{equation}
\begin{equation} \label{4}
\frac{\partial \Psi }{\partial r_{12}} \mid_{r_{12}=0}
=\frac{1}{2} \Psi(R,R,0).\ \ [r_{1}=r_{2}=R]
\end{equation}
The conditions (2) and (3) pertain to the situation when the
coordinates of one of the electrons and the nucleus coinside. These
conditions are valid for the electrons, which have the same (triplet
states) or the opposite (singlet states) spin directions. The
condition (4) deals with coincidence of coordinates of two
electrons. It is valid only for the singlet states, while due to the
Pauli exclusion principle \( \Psi (R,R,0)=0 \) for the triplet
states. The inclusion of functions with such cusps into trial wave
functions has been shown to improve dramatically the rates of
convergence of Rayleigh-Ritz variational calculations \cite{4}. The
using of the proper correlation function, which obey the Kato
conditions (2)-(4), accelerates considerably the convergence of
CFHHM \cite{2} approach, as well.

It is known that the cusp conditions (2)-(4) are consequences of the
Coulomb singularity in the potential and provide specific relations
between the wave function and its first derivative at the points of
coalescence. It was shown in Ref. [5] that the coalescence behaviour
also uniquely determines the third derivative of the spherically
averaged wave function in terms of the lower derivatives. The
deduced relations are valid for any atom, molecule, or electron gas
in any smooth external field.

There are also singularities involving more than two particle, such
as the triple-coincidence singularity in the Helium atom, when both
electrons simultaneously approach the nucleus. A formal expansion in
powers of the hyperradius \( r=\sqrt{r^{2}_{1}+r^{2}_{2}} \) and its
logarithm \( \ln r \) about this singular point was proposed by Fock
\cite{6} for the \( S \)-state wave functions of the Helium atom.
Subsequently, much efforts has been devoted to understanding this
expansion. The \( O(r^{0}) \), \( O(r^{1}) \) , and \( O(r^{2}\ln r)
\) terms in Fock's expansion are easy to obtain analytically. The \(
O(r^{2}) \) term in the expansion has been obtained in closed form
by Maslen and co-workers, through the extensive use of computer
algebra \cite{7,8}. Myers and co-authors \cite{9} have examined
their results, and have verified that the inclusion of this term in
the expansion yields a continuous "local" energy, whereas the
"local" energy is finite but discontinuous at \( r=0 \) if the term
is omitted. Forrey \cite{10} performed variational calculations of
the ground-state energy of Helium. His basis set included an
explicit treatment of the Fock expansion in hyperspherical
coordinates and extra products of Laguerre polynomials in perimetric
coordinates. This work has demonstrated that the use of Fock basis
functions provided a substantial improvement in the convergence
rate.

We would like to emphasize that the calculation of the accurate wave
function at the coalescence lines is a very difficult problem just
because of their cusp nature. On the other hand, a number of atomic
physics problems could be solved by using the functions appearing on
the RHS of Eqs.\ (2)-(4). The processes of photoionization in the
Helium atom and heliumlike ions \cite{11} could serve as an example.

It is well-known (see, e. g.,\cite{7,8,9}), that using Hamiltonian
(1) we can present the Schr\"{o}dinger equation for \( S \)-states
of two-electron atom/ions in the form
%\begin{equation} \label{5}
%\begin{array}{c}
\begin{eqnarray}
&&-\frac{1}{2}
\left [
\frac{\partial ^{2}\Psi }{\partial r^{2}_{1}}+\frac{\partial ^{2}
\Psi }{\partial r^{2}_{2}} +2\frac{\partial ^{2}\Psi } {\partial
r^{2}_{12}}+\frac{2}{r_{1}}\frac{\partial \Psi }{\partial r_{1}}
+\frac{2}{r_{2}}\frac{\partial \Psi }{\partial
r_{2}}+\frac{4}{r_{12}}\frac{\partial \Psi }{\partial r_{12}}
\right .
\\
%~~~~~~~~~~~~~~~~~~~~~~~~~~\\
&&\left .
+\left(\frac{r^{2}_{1}-r^{2}_{2}+r^{2}_{12}}{r_{1}r_{12}}\right)
\frac{\partial ^{2}\Psi }{\partial r_{1}\partial r_{12}}
+\left(\frac{r^{2}_{2}-r^{2}_{1}+r^{2}_{12}}{r_{2}r_{12}}\right)
\frac{\partial ^{2}\Psi }{\partial r_{2}\partial r_{12}}
\right ]
 =
\left( \frac{Z}{r_{1}}+\frac{Z}{r_{2}}-\frac{1} {r_{12}}+ E
\right)\Psi. \nonumber
\end{eqnarray}
%\end{equation}
In this paper we provide the accurate analytic solutions of the Schr\"{o}dinger
equation\ (5) at the coalescence lines for both small and very large \( R \).
The Kato cusp conditions (2)-(4) are employed to solve the problem.

\section{Electron-nucleus coalescence}

To investigate the case of coalescence of one electron and the
nucleus in two-electron atom/ions, one should find the limit as,
e.g., \( r_{2} \) approaches zero for both sides of Eq.\ (5). It is
easier to perform this mathematical operation with the help of
following relations:
\begin{equation}\label{6}
\frac{r^{2}_{1}-r^{2}_{2}+r^{2}_{12}}{2r_{1}r_{12}}=\cos \theta _{1};
~~~~~~~~~~~~~\frac{r^{2}_{2}-r^{2}_{1}+r^{2}_{12}}{2r_{2}r_{12}}=\cos \theta _{2},
\end{equation}
where \( \theta _{1} \) is the angle between the vectors \( \mathbf{r}_{1} \)
and \( \mathbf{r}_{12} \), and \( \theta _{2} \) is the angle between
\( \mathbf{r}_{2} \) and \( \mathbf{r}_{12} \) (see Fig.\ 1). It is
clear that:
\begin{equation}\label{7}
\lim _{_{r_{2}\rightarrow 0}}\theta _{1}=0;
~~~~~~~~~~~~~~~~~~\lim _{_{r_{2}\rightarrow 0}}\theta _{2}=\pi /2.
\end{equation}

%\begin{figure}
%\begin{center}
%\{fig1.eps}}
%\end{center}
%\caption{Interparticle coordinates and associated angles for the two-electron
%atom/ions.}
%\end{figure}

\begin{figure}
\begin{center}
\epsfig{file=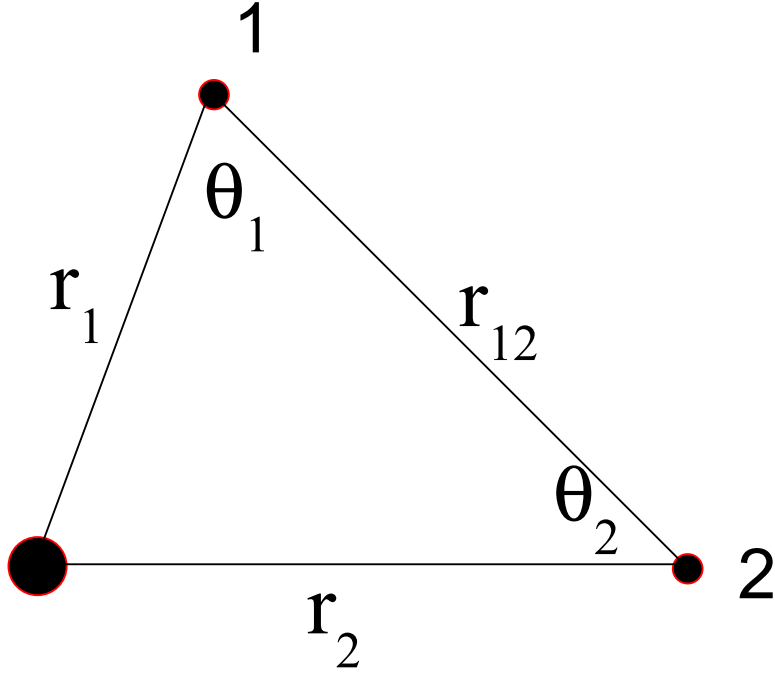,width=120mm}
\end{center}
\caption{Interparticle coordinates and associated angles for the
two-electron atom/ions.}
\label{fig1}
\end{figure}

Then, using Eqs.\ (6), (7), we can rewrite Eq.\ (5), taking the limit as \( r_{2} \)
approaches zero:
%\begin{equation}\label{8}
%\begin{array}{c}
\begin{eqnarray}
&&-\frac{1}{2}
{\left[\frac{\partial ^{2}\Psi }{\partial r^{2}_{1}} +
\frac{\partial ^{2}\Psi }{\partial r^{2}_{2}} + 2
{\left( \frac{\partial ^{2}\Psi }{\partial r^{2}_{12}}+
\frac{\partial ^{2}\Psi }{\partial r_{1}\partial r_{12}}\right)}
\right]}{\mid }
_{_{r_{2}=0}} - \frac{1}{R}
{\left( \frac{\partial \Psi }{\partial r_{1}} + 2
\frac{\partial \Psi }{\partial r_{12}}\right)}
{\mid }_{_{r_{2}=0}}
~\nonumber\\
&&=\frac{1}{r_{2}}
{\left( \frac{\partial \Psi }{\partial r_{2}} + Z\Psi\right)}
{\mid}_{_{r_{2}=0}} +
{\left( \frac{Z-1}{R} + E \right)}\Psi (R,0,R).%~~~~~~~~~~~
%~~~~~~~~~~~\\
\end{eqnarray}
%\end{array}
%\end{equation}
Here we took into consideration that \( r_{1}=r_{12}=R \) as \( r_{2} \) approaches
zero. Inserting the Kato condition (3) into the RHS of Eq.\ (8), and assuming that
the terms of Eq.\ (8) must be finite in the whole 3-D space, we obtain
%\begin{equation}\label{9}
%\begin{array}{c}
\begin{eqnarray}
{\left[
\frac{\partial ^{2}\Psi }{\partial r^{2}_{1}} +
\frac{\partial ^{2}\Psi }{\partial r^{2}_{2}} + 2
\left(
\frac{\partial ^{2}\Psi }{\partial r^{2}_{12}} +
\frac{\partial ^{2}\Psi }{\partial r_{1}\partial r_{12}}
\right)
\right] }
\mid_{_{r_{2}=0}} +
\frac{2}{R}
{\left(
\frac{\partial \Psi }{\partial r_{1}} + 2
\frac{\partial \Psi }{\partial r_{12}}
\right)}
\mid_{_{r_{2}=0}}
~\nonumber \\
= 2 \left(
\frac{1-Z}{R} - E
\right)
\Psi (R,0,R).%~~~~~~~~~~~~~~~~~~~~~~~~~~~~~~~~~~~~~~~~~~~~
\end{eqnarray}
%\end{array}
%\end{equation}
We could certainly obtain the same equation if took the limit as \(
r_{1} \) approaches zero (\( r_{2}=r_{12}=R \)).

It is seen that the LHS of Eq.\ (9) presents a sum of the form
\( \sum _{klm}c_{klm}(R)\Lambda _{lm}^{(k)}(R) \),~~where for the partial derivatives
of \( \Psi  \) taken at the electron-nucleus coalescence line, we have
\begin{equation}\label{10}
\Lambda _{l,m}^{(k)}(R) = \lim _{_{r_{2}\rightarrow 0}}
\frac{\partial ^{k}\Psi (r_{1},r_{2},r_{12})}{\partial r_{l}
\partial r_{m}}\ \ \ ( k=1,2;~~~l,m=0,1,2,12).
%\partial r_{m}}~~~~~~~~~~~~~~~~~~~~~~~~~~~~~~( k=1,2;~~~l,m=0,1,2,12).
\end{equation}
Here \( \partial r_{0}=1 \). Then, in the case of the first partial derivatives
we have \( k=1 \), whereas one of the numbers \( l,m \) is equal to zero.

Let us now denote the two-electron wave \( S \)-function at the electron-nucleus
coalescence line as
\begin{equation}\label{11}
\Psi (R,0,R)\equiv F(R).%~~~~~~~~~~~~~~~~~~~~~~~~~~~~~~~~~~~~~~~~~~~~~~~~~~~~~
\end{equation}
So, if we could express all of the functions \( \Lambda _{l,m}^{(k)}(R) \)
{\large } through the functions \( F(R) \), \( F^{\prime }(R) \) and \( F^{\prime \prime }(R) \),
with factors being depending on \( R \), then we obtain an ordinary
differential equation of the second order for the function \( F(R) \). The
prime denotes differentiation, as usual.

Solution of this differential equation under the proper boundary
conditions could give us the desired function \( F(R) \). We do not
yet know how to do this in general form.
%Perhaps, we or somebody else will solve this difficult problem in future.
However, as a first but important step we propose here a method for
solving Eq.\ (9) in the boundary regions, i.e., at very large \( R
\) and at small \( R \). One should notice that the numerical
calculation of \( F(R) \) in these regions is particularly
difficult.

The direct Correlation Function Hyperspherical Harmonic Method allows to calculate
numerically the two-electron wave function \( \Psi (r_{1,}r_{2,}r_{12}) \), as well as
its special case \( F(R) \), with very large accuracy. By using
the CFHHM numerical calculations we obtained the following limit relations between
the functions mentioned above for the asymptotic region of very large \( R \):
\begin{equation}\label{12}
\lim _{_{R\rightarrow \infty }} \left\{ \frac{\Lambda _{1,0}^{(1)}(R)}{F^{\prime }(R)}\right\} = 1,
\end{equation}
\begin{equation}\label{13}
\lim _{_{R\rightarrow \infty }} \left\{ \frac{\Lambda _{1,1}^{(2)}(R)}{F^{\prime \prime }(R)}\right\} = 1,
\end{equation}
\begin{equation}\label{14}
\lim _{_{R\rightarrow \infty }} \left\{ \frac{\Lambda _{2,2}^{(2)}(R)}{F(R)}\right\} = Z^{2},
\end{equation}
\begin{equation}\label{15}
\lim _{_{R\rightarrow \infty }} \left\{ \frac{\Lambda _{1,12}^{(2)}(R)}{F(R)}\right\} = 0,
\end{equation}
\begin{equation}\label{16}
\lim _{_{R\rightarrow \infty }} \left\{ \frac{\Lambda _{12,12}^{(2)}(R)}{F(R)}\right\}= 0,
\end{equation}
\begin{equation}\label{17}
\lim _{_{R\rightarrow \infty }} \left\{ \frac{\Lambda _{12,0}^{(1)}(R)}{F(R)}\right\} = 0.
\end{equation}
The calculations show that these relationships are valid at least to
four significant digits. We cannot achieve higher accuracy due to
the fact that the inaccuracies of the wave functions and especially
of their derivatives go up with \( R \). Note that the asymptotic
relations (12)-(17) are valid for the ground states of the
two-electron atom/ions, as well as for its \emph{excited} states,
including \emph{triplet} states. Of course, we were not able to
perform the proper calculations for all of the excited states.
However, we could guarantee the validity of Eqs.\ (12)-(17) for
several first ones.

Relations (15)-(17) show that we can neglect the partial derivatives with respect to
\( r_{12} \). Moreover, the calculations of the accurate CFHHM wave functions
show that the ratio of \( F^{\prime }(R)/F(R) \) achieves a finite value
as \( R \) approaches infinity. In the next sections we will obtain this ratio
as a finite function of \( Z \) and \( E \). This property together with the
limit relation (12) allows us to neglect the terms proportional to \( 1/R \)
on both sides of Eq.\ (9). And finally, using
the relations (13), (14), Eq.\ (9) is transformed into the following simple
differential equation:
\begin{equation}\label{18}
\frac{d^{2}F_{as}(R)}{dR^{2}}+(2E+Z^{2})F_{as}(R)=0.
\end{equation}
As is well known, the solution, which is convergent at \( R\rightarrow \infty  \),
has the form
\begin{equation}\label{19}
F_{as}(R) = C_{1}\exp \left( -R\sqrt{-2E-Z^{2}}\right).
\end{equation}
The function \( F_{as}(R) \) is the asymptotic representation (for
very large \( R \)) of the accurate two-electron wave function in
the situation when one of the electrons "seats" on the nucleus,
while the other electron is far away. Eq.\ (12) shows that this
function depends on two parameters, the nuclear charge \( Z \) and
the total energy \( E \) of the two-electron atomic system. \( C_{1}
\) is an arbitrary constant. We used only the accurate wave
functions of the discrete spectrum (\( E<0 \)) to obtain the
relations (12)-(17). The condition of exponent in Eq.\ (19) to be
real leads to the inequality
\begin{equation}\label{20}
-2E>Z^{2}
\end{equation}
at least for the \( S \)-states of the Helium atom or heliumlike ions with the
nuclear charge \( Z \).

We have calculated the CFHHM energy levels of the Helium atom and the ions of
Li\( ^{+} \)and B\( ^{3+} \) for both the singlet and triplet \( S \)-states with
\( n\leq 7 \). These data, presented in Table I, confirm the
validity of the inequality \ (20).
\begin{table}[hbt]
\begin{center}
\caption{Energy levels (in a. u.) of \( n^{k}S \) states for the Helium
atom and heliumlike ions of \( Li^{+}(Z=3) \) and \( B^{3+}(Z=5) \).}
\begin{tabular}{|c|c|c|c|}
\hline
\( n \)~\textbackslash{}&
\( He: \) ~ \( k=1 \) ~~~~\( k=3 \)~~~~~~&
\( Li^{+}: \) ~ \( k=1 \)~~~~\( k=3 \)~~~~~~&
\( B^{3+}: \) ~ \( k=1 \)~~~~\( k=3 \)~~~~~\\
\hline
\hline
1&
~-2.903724~~~~~~~~~~~~~~~~~~~&
~-7.278876~~~~~~~~~~~~~~~~~~~&
~-22.02788~~~~~~~~~~~~~~~~~~\\
\hline
2&
-2.145970~~~~-2.175225&
-5.040179~~~~-5.110019&
-14.57652~~~-14.73188\\
\hline
3&
-2.061221~~~-2.068696&
-4.733102~~~-4.751430&
-13.41017~~~-13.45127\\
\hline
4&
-2.033566~~~-2.036524&
-4.629208~~~-4.636571&
-13.00797~~~-13.02461\\
\hline
5&
-2.021225~~~-2.022633&
-4.581895~~~-4.585572&
-12.82326~~~-12.83163\\
\hline
6&
-2.014537~~~-2.015122&
-4.556331~~~-4.558559&
-12.72341~~~-12.72824\\
\hline
7&
-2.010629~~~-2.010870&
-4.541111~~~-4.542445&
-12.66341~~~-12.66654\\
\hline
\end{tabular}
\end{center}
\end{table}

It is easy to conclude that Eq.\ (20) corresponds to the first
ionization threshold \( I_{1}=Z^{2}/2 \), and that the limit
relations (12)-(17) describe the ground and \emph{singly} excited
bound states {[}12,13{]}. So, if the electrons are far away from
each other (\( R\rightarrow \infty  \)), then the simplest model one
may think of is the model of two independent electrons. The inner
electron is bound in a state with principal quantum number \( N \)
and energy \( E_{N}=-Z^{2}/(2N^{2}) \), the outer electron is in a
hydrogen-like orbital with energy \( E_{n}=-(Z-1)^{2}/(2n^{2}) \)
and \( n\geq N \) assuming a screening of the nuclear charge. The
total energy is simply the sum of the one-particle energies. The
case of \( R \) approaching infinity corresponds to the conditions
of \( n\rightarrow \infty  \), \( E_{n}\rightarrow 0 \), and \(
E_{thresh}\equiv I_{1}=E_{N=1} \). So, these arguments give
additional evidence to the validity of the limit relations (12)-(17)
and, consequently, the asymptotic solution (19) for the ground and
\emph{singly} excited bound states.

We have obtained some limit relations for the functions (10) and
(11) in the vicinity of the triple collision point, \( R=0 \), as
well. The proper numerical calculations yield for the singlet
states:
\begin{equation}\label{21}
\lim _{_{R\rightarrow 0}}\left\{\frac{\Lambda _{1,0}^{(1)}(R)}{F(R)}\right\} = -Z,
\end{equation}
\begin{equation}\label{22}
\lim _{_{R\rightarrow 0}}\left\{\frac{\Lambda _{12,0}^{(1)}(R)}{F(R)}\right\} =\frac{1}{2},
\end{equation}
\begin{equation}\label{23}
\lim _{_{R\rightarrow 0}}\left\{\frac{\Lambda _{1,12}^{(2)}(R)}{F(R)}\right\} =-\frac{1}{2}Z.
\end{equation}
Note that Eq.\ (21) contains the singlet function \( F(R) \), whereas the corresponding
Eq.\ (12) includes the first derivative \( F^{\prime }(R) \). The results for
the triplet states are:
\begin{equation}\label{24}
\lim _{_{R\rightarrow 0}}\left\{ \frac{R^{2}\Lambda _{1,1}^{(2)}(R)}{F(R)}\right\}
 = \lim _{_{R\rightarrow 0}}\left\{ \frac{R\Lambda _{1,0}^{(1)}(R)}{F(R)}\right\} = 2,
\end{equation}
\begin{equation}\label{25}
\lim _{_{R\rightarrow 0}}\left\{ \frac{R^{2}\Lambda _{2,2}^{(2)}(R)}{F(R)}\right\} = -2,
\end{equation}
\begin{equation}\label{26}
\lim _{_{R\rightarrow 0}}\left\{ \frac{R\Lambda _{1,12}^{(2)}(R)}{F(R)}\right\} = \frac{1}{2},
\end{equation}
\begin{equation}\label{27}
\lim _{_{R\rightarrow 0}}\left\{ \frac{\Lambda _{12,0}^{(1)}(R)}{F(R)}\right\} = \frac{1}{4},
\end{equation}
As is known, the two-electron wave functions of the singlet states
at the triple coincidence point are non-zero. Therefore, Eqs.\ (21),
(22) allows to avoid the divergence at \( R=0 \) for terms
proportional to \( R^{-1} \) in Eq.\ (9). The triplet states, which
are proportional to \( R^{2} \) in the vicinity of the triple
collision point, don't have to obey such a requirement.

We were able to obtain only the simplest limit relations as \( R \)
approaches zero. We hope these relations could be good for searching
the general solution of Eq.\ (9). However, to derive the solution
for small \( R \) we propose another way, which is more precise as
well as more reliable. As was mentioned earlier, in Refs.\cite {7,8}
analytic expansions of the three-body atomic wavefunctions were
presented. The expansions were derived for the exact solutions of
the Schr\"{o}dinger equation\ (5) up to the terms of the order \(
r^{2} \) (including \( r^{2}\ln r \)). Note that limit relations (21)-(27)
(\(R\rightarrow 0 \)) could be obtained by using Ref.\ [8] too.
We used some of those results (see Ref.\ [8], pp. 2796-2797) to obtain the analytical
representation of the two-electron wave functions at the
two-particle coalescence lines in the vicinity of the triple
coincidence point. The same results, but for the singlet states
only, could be derived by using Ref.\ [9]. However, one should be
very careful, because in the last reference we found at least three
misprints, which could have influence on the final results. The
first misprint is a missing factor 2 in the expression for \(
Y_{2,0} \) (below Eq.\ (14) \cite{9}). The second one is the
incorrect expression \( r_{12}\sin \alpha \cos \theta  \) on the RHS
of the expression for \( Y_{2,1} \) (below Eq.\ (14) \cite{9}). And
the third misprint is the missing function \( \cos ^{-1} \) before
\( (\mathbf{r}_{1}\cdot \mathbf{r}_{2}/r_{1}r_{2}) \) in the RHS of
Eq.\ (11) \cite{9}.

So, using the results of Refs.\ [8,9] and taking the limit as
\(r_{2}\rightarrow 0 \), we obtain for the singlet states:
\begin{equation}\label{28}
F(R)\simeq 1-R\left( Z-\frac{1}{2}\right) + \frac{R^{2}}{12}[4Z^{2}-2Z(3-\ln 2)+1-2E].
\end{equation}
The similar result for the triplet states has the form
\begin{equation}\label{29}
F(R)\simeq R^{2} \left \{1-R\left( \frac{2}{3}Z-\frac{1}{4}\right) +
\frac{R^{2}}{10}{\left [ \frac{5}{3}Z^{2}-Z(2-\frac{5}{6}\ln 2)+\frac{1}{4}-E\right ]}\right\}.
\end{equation}
For simplicity, the wave functions (28), (29) are normalized by
condition \( F(0)=1 \) for the singlet states, and \( \left[
F(R)/R^{2}\right] _{R=0}=1 \) for the triplet states.

We would like to pay particular attention to the \emph{absence of the Fock's
logarithmic term} in both expressions. This term disappears at the electron-nucleus
coalescence line, because of the vanishing Fock's angular coefficient \( \psi _{21} \) in
the limit as \( r_{2} \) approaches zero (whereas \( r_{1}\rightarrow r_{12} \))
for the singlet states.

At first glance, it is natural to assume that all of the Fock's
logarithmic terms are canceled at the electron-nucleus coalescence
line. However, such an assumption proved to be incorrect. We
verified the angular coefficients \( \psi _{31} \), \( \psi _{41} \)
and \( \psi _{42} \), corresponding to the logarithmic terms in the
Fock's expansion up to the terms of order \( r^{4} \), \( r^{4}\ln
^{2}r \) (singlet states). The exact expressions for these
quantities could be found, e.g., in Ref.\ [7]. All of these three
angular coefficients proved to be nonzero. So, we conclude that the
first logarithmic term \( \psi _{21}r^{2}\ln r \) of the Fock's
expansion is the only one to vanish at the electron-nucleus
coalescence line, at least for the singlet states. Eq.\ (29) shows
additionally that for the triplet states all of the logarithmic
terms, up to the fourth order in \( r \), are canceled in the limit
as \( r_{2}\rightarrow 0 \) (or \( r_{1}\rightarrow 0 \)).
Accordingly, the values of the first and second derivatives \(
F^{\prime }(0) \), \( F^{\prime \prime }(0) \) for the singlet
states and \( \left[ F(R)/R^{2}\right] ^{\prime }_{R=0} \), \(
\left[ F(R)/R^{2}\right] ^{\prime \prime }_{R=0} \) for the triplet
states are finite. We verified the validity of the expansions (28),
(29) by direct calculation of these derivatives in the limit as \(
R\rightarrow 0 \), using the accurate \( \Psi \)-functions. The
results coincided with the calculations performed according to the
analytical formulas (28), (29) within the accuracy of five
significant digits.

\section{Electron-electron coalescence}

For the case of forming the two-electron coalescence or the
coincidence of the coordinates of the both electrons, one should
take the limit as \(r_{12}\rightarrow 0 \) on both
sides of the Eq.\ (5). In this case we have (see Fig.\ 1):
\begin{equation}\label{30}
\lim _{_{r_{12}\rightarrow 0}}\theta _{1} = \lim _{_{r_{12}\rightarrow 0}}\theta _{2} = \pi /2.
\end{equation}
Then, in the limit as \( r_{12}\rightarrow 0 \) both terms with the mixed
partial derivatives vanish in Eq.\ (5), and we can write:
%\begin{equation}\label{31}
%\begin{array}{c}
\begin{eqnarray}
&&-\frac{1}{2} \left( 2 \frac{\partial ^{2}\Psi }{\partial r^{2}_{1}}
+ 2 \frac{\partial ^{2}\Psi }{\partial r^{2}_{12}} \right) \mid _{_{r_{12}=0}}
- \frac{2}{R} \frac{\partial \Psi }{\partial r_{1}} \mid_{_{r_{12}=0}} \nonumber\\
%~~~~~~~~~~~~~~~~~~~~~~~~~~~\\
&&= \frac{2}{r_{12}} \left( \frac{\partial \Psi }{\partial r_{12}} - \frac{1}{2}\Psi \right)
\mid_{_{r_{12}=0}} + \left( \frac{2Z}{R}+E\right) \Psi (R,R,0).%~~~~~~~~~~~~~~~~
\end{eqnarray}
%\end{array}
%\end{equation}
When deriving Eq.\ (31), we took into account that \( r_{1}=r_{2}=R \) as
\( r_{12} \) approaches zero. Then, using the Kato condition (4) in the RHS
of Eq.\ (31), we obtain
\begin{equation}\label{32}
\left( \frac{\partial ^{2}\Psi }{\partial r^{2}_{1}} + \frac{\partial ^{2}\Psi }{\partial r^{2}_{12}}
\right) \mid_{_{r_{12}=0}} + \frac{2}{R} \frac{\partial \Psi }{\partial r_{1}} \mid_{_{r_{12}=0}}
 = -\left( \frac{2Z}{R}+E\right) \Psi (R,R,0).
\end{equation}
From the CFHHM numerical calculations, we obtain the following relation
\begin{equation}\label{33}
\lim _{_{r_{12}\rightarrow 0}}\frac{\partial \Psi (r_{1},r_{2},r_{12})}{\partial r_{1}}
 = \frac{1}{2} \frac{d\Phi (R)}{dR},
\end{equation}
where
\begin{equation}\label{34}
\Phi (R) =  \Psi (R,R,0).
\end{equation}
The relation (33) is not a double-limit relation like all the other similar relations
presented in the Section II. It is the only relation of such form obtained.
It is valid for all \( 0\leq R\leq \infty  \) and \( Z\geq 1 \). The excited
states are included as well. Using Eq.\ (33), we can rewrite Eq.\ (32) in the
form
\begin{equation}\label{35}
\left( \frac{\partial ^{2}\Psi }{\partial r^{2}_{1}} +
\frac{\partial ^{2}\Psi }{\partial r^{2}_{12}} \right)
\mid_{_{r_{12}=0}} = g(R),
\end{equation}
where
\begin{equation}\label{36}
g(R) = -\frac{1}{R} \frac{d\Phi (R)}{dR} -
\left( \frac{2Z}{R} + E \right)\Phi (R).
\end{equation}
Calculation of the ordinary first derivative \( d\Phi /dR \) is considerably
more precise than the calculation of the partial derivatives of the second order.
The proper calculation of the function \( g(R) \) and its comparison with the
ordinary second derivative \( d^{2}\Phi /dR^{2} \) yields, in the limit of \( R \)
approaching infinity:
\begin{equation}\label{37}
\lim _{_{R\rightarrow \infty }} \left\{ \frac{\Phi ^{\prime \prime }(R)}{g(R)}\right\} = 4.
\end{equation}
Then, using the limit relation (37), we can rewrite Eq.\ (36) in the limit of
very large \( R \):
\begin{equation}\label{38}
\frac{1}{4}\frac{d^{2}\Phi _{as}}{dR^{2}} + E\Phi _{as} = 0.
\end{equation}
Like before, we neglected the terms proportional to \( R^{-1} \) in Eq.\ (36).
The proper solution of the differential equation (38) has the form
\begin{equation}\label{39}
\Phi _{as}(R)=\overline{C}_{1}\exp \left( -2R\sqrt{-E}\right),
\end{equation}
where \( \overline{C}_{1} \) is an arbitrary constant. The function \( \Phi _{as}(R) \)
is the asymptotic representation of the accurate two-electron atomic wave function
for the case of two electrons being at the same point but far away from the
nucleus. Unlike \( F_{as} \), Eq.\ (39) shows that \( \Phi _{as} \) depends on only one parameter,
energy \( E \). As far as we consider the
discrete spectrum, then \( E<0 \) and consequently the exponent
on the RHS of Eq.\ (39) is negative.

Moreover, we obtained the following relation in the limit as \( R\rightarrow 0 \):
\begin{equation}\label{40}
\lim _{_{r_{1}=r_{2}=R\rightarrow 0}}\lim _{_{r_{12}\rightarrow 0}}
\frac{\partial \Psi (r_{1},r_{2},r_{12})}{\partial r_{1}} =
-Z\lim _{_{R\rightarrow 0}}\Phi (R).
\end{equation}
It enables one to avoid the divergence at \( R=0 \) for the terms that are proportional
to \( R^{-1} \) in the general Eq.\ (32) for the case, when two electrons are at the same point.

The approximate solution of the Eq.\ (32) at small \( R \) may now be obtained
by neglecting the terms that are not proportional
to \( R^{-1} \) in that equation. Using also Eq.\ (33), we have
the equation
\begin{equation}\label{41}
\frac{d\Phi _{2}}{dR} + 2Z\Phi _{2} = 0,
\end{equation}
with the the solution
\begin{equation}\label{42}
\Phi _{2}(R) = \overline{C}_{2}\exp (-2ZR).
\end{equation}
We can again use the approach developed in the works \cite{8,9} at 
\( r_{12}\rightarrow 0 \) (\( r_{1}=r_{2}=R \)). However, one shouldn't
forget that the expressions presented in these papers are valid only for \( r_{2}<r_{1} \).
In Ref.\cite{10} the angular coefficients \( \psi _{00},\psi _{10},\psi _{21},\psi _{20} \)
of the Fock expansion are expressed through Pluvinage coordinates \( \zeta  \)
and \( \eta  \) \cite{14}, which are more symmetric and are valid for arbitrary
\( r_{1} \) and \( r_{2} \). Taking into account that the scaling transformation
\( \mathbf{r}\rightarrow \mathbf{r}/Z \) was applied to the Hamiltonian in
Ref.\ [14], we obtained under the simplifying normalizing condition of \( \Phi (0)=1 \):
\begin{equation}\label{43}
\Phi (R)\simeq 1-2ZR + R^{2} \left[ \frac{2Z}{3}\left( \frac{2}{\pi }-1\right)
\ln R - \frac{E}{3} + \frac{5}{3} Z^{2} + \delta _{z} \right ]
\end{equation}
with
\begin{equation}\label{44}
\delta _{z} = \frac{2Z}{3}\left( \frac{1}{\pi }-1\right)\ln 2 +
\frac{1}{6} + \frac{Z}{3} + 2a_{2,1}.
\end{equation}
Here \( a_{2,1} \) is an unknown coefficient of the homogeneous solution, which
cannot be deduced from the local behaviour of the Schr\"{o}dinger equation near
the nucleus \cite{8,9,10}. It is a very difficult problem to calculate exactly \( a_{2,1} \),
because of the presence of the logarithmic term on the RHS of Eq.\ (43). 
However, we have evaluated \( \delta _{z} \) (and consequently \( a_{2,1} \))
using the accurate CFHHM wave functions. This yielded the following approximate
dependence:
\begin{equation}\label{45}
\delta _{z}\simeq \frac{1}{2} - \frac{3}{10} Z.
\end{equation}
This linear part of the \( Z \)-dependence is the most accurate one for \( 2\leq Z\leq 5 \).
We found that the term \( -E/3 \) in (43) expresses correctly the dependence of
\( \Phi (R) \) on the state of excitation, while the term \( (5/3)Z^{2} \)
represents accurately the nonlinear dependence of \( \Phi (R) \) upon the nuclear
charge \( Z \). Note that using the proper formulas from Refs.\ [8,9] yields
the erroneous term \( Z^{2} \) instead of the correct result \( (5/3)Z^{2} \)
in the expression (43). As we can see, the Fock's logarithmic term is preserved at
the electron-electron coalescence line, unlike the case of the electron-nucleus
line. Hence, the second derivative \( \Phi ^{\prime \prime }(R) \) has a logarithmic
singularity at the origin. It is seen that solutions (42) and (43) coincide in 
the first order approximation at small \( R \).

\section{Results and discussions}

In Sections II, III we have obtained analytical representations of the accurate
two-electron wave functions at the boundary regions of both two-particle
coalescence lines. We found that the behaviour of these boundary solutions
both at small and at large distances \( R \) has an exponential character.
These properties enable us to propose a simple approximate representation for the
ground state wave functions in the two-exponential form. The main idea is that
the first exponential represents the behaviour at small \( R \) and the second
exponential represents the behaviour at very large \( R \). So, let us consider the following function:
\begin{equation}\label{46}
f(R) = C\left\{ \exp (-\lambda R)+\gamma \exp (-\beta R)\right\},
\end{equation}
with
\begin{equation}\label{47}
\lambda =\alpha (1+\gamma )-\gamma \beta.
\end{equation}
Here \( C,\, \alpha ,\, \beta  \) and \( \gamma  \) are arbitrary parameters.
This two-exponential function possesses an important peculiarity:
\begin{equation}\label{48}
\frac{f^{\prime }(0)}{f(0)} = -\alpha,
\end{equation}
that is the ratio of the first derivative to the function itself at the origin
(\( R=0 \)) depends upon only one parameter \( \alpha  \). Let the second
exponential in the RHS of Eq.\ (46) present the behaviour of the wave functions
in question at very large \( R \). Then according to the results of the previous 
sections (see Eqs.\ (19), (28)), we can put \( \beta _{s}=\sqrt{-2E-Z^{2}} \)
and \( \alpha _{s}=Z-1/2 \) for the electron-nucleus coalescence line of the
singlet states. Considering \( C \) as the normalization constant, we have
only one unknown parameter, \( \gamma  \). It can be obtained using the
second derivative of the wave function \( F(R) \) at the origin. Then
the double differentiation of the general function (46) yields, in the limit
of \( R \) approaching zero,
\begin{equation}\label{49}
h\equiv \frac{f^{\prime \prime }(0)}{f(0)} =
\frac{[\alpha (1+\gamma )-\gamma \beta ]^{2}+\gamma \beta ^{2}}{1+\gamma }.
\end{equation}
On the other hand, we have from Eq.\ (28):
\begin{equation}\label{50}
h_{s} = \frac{1}{6}[4Z^{2}-2Z(3-\ln 2)+1-2E].
\end{equation}
Eq.\ (49) is a quadratic with respect to parameter \( \gamma  \). It
has two roots: \( \gamma _{1}=-1 \) and \( \gamma _{2}=(h-\alpha ^{2})/(\alpha -\beta )^{2} \).
While the root \( \gamma _{1} \) yields a trivial solution \( f(R)=0 \), we
retain only the solution \( \gamma =\gamma _{2} \). Using Eq.\ (50)
and taking into consideration the values of parameters \( \alpha _{s} \) and
\( \beta _{s} \) mentioned above, we obtain
\begin{equation}\label{51}
\gamma _{s}= \frac{4(Z\ln 2-E-Z^{2})-1}{3\left( 1-2Z+2\sqrt{-2E-Z^{2}}\right) ^{2}}
\end{equation}
The exponent \( \lambda  \) of the first exponential plays an important part
in constructing the approximate function (46). Note that it does not represent
the behaviour of this function at the origin. Both exponentials
give contributions to the correct behaviour of the function (46) in this boundary
region according to Eqs.\ (48), (49). However, the inequality \( \lambda >\beta  \)
must be valid if we want the exponent \( \beta  \) to present the approximate
function (46) at very large \( R \). So, using Eq.\ (51) and the proper expressions
for parameters \( \alpha _{s} \) and \( \beta _{s} \) we have for the ground
state
\begin{equation}\label{52}
\lambda _{s} = \frac{2-4E+8Z^{2}-4Z[3(1+\beta _{s})-\ln 2]+6\beta _{s}}{6[2(Z-\beta _{s})-1]}.
\end{equation}
In Table II we present the numerical values of the exponents \( \lambda _{s} \)
and \( \beta _{s} \) as well as the factor \( \gamma _{s} \) for the Helium atom
and several heliumlike ions.
\begin{table}[hbt]
\begin{center}
\caption{The parameters \( \lambda  \), \( \beta  \) and \( \gamma  \)
for the approximate wave functions at the electron-nucleus coalescence line.
The lower index \( s \) signifies that the parameter presents a singlet state.}
\begin{tabular}{|c|c|c|c|c|}
\hline
\( Z \)&
2&
3&
4&
5\\
\hline
\hline
\( \lambda _{s} \)&
1.58574&
2.75657&
3.93411&
5.11350\\
\hline
\( \beta _{s} \)&
1.34441&
2.35793&
3.36320&
4.36599\\
\hline
\( \gamma _{s} \)&
0.551062&
1.80594&
3.17337&
4.57790\\
\hline
\end{tabular}
\end{center}
\end{table}
As one can see from Table II, the parameter \( \lambda _{s} \) is greater than
\( \beta _{s} \) for all \( Z \) presented.

\begin{figure}
\begin{center}
\epsfig{file=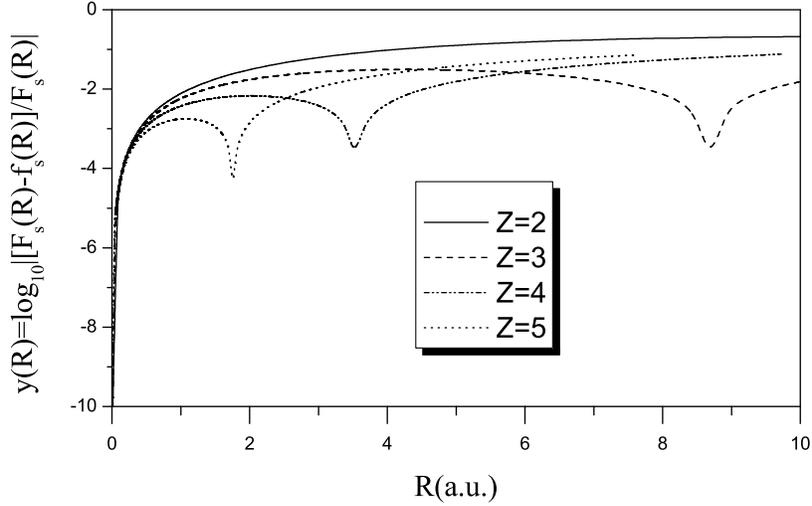,width=120mm}
\end{center}
\caption{Deviation of the approximate function \( f_{s}(R) \) from
the exact value \( F_{s}(R) \) at the electron-nucleus coalescence
line for the ground states of the two-electron atom/ions with the
nuclear charge \( Z=2,3,4,5 \).}
\label{fig2}
\end{figure}

To verify how good are the approximate functions (46) with the parameters of
Table II, we have drawn the curves \( y(R)=\log _{10}\left| \frac{F_{s}(R)-f_{s}(R)}{F_{s}(R)}\right|  \)
on Fig.\ 2. \( F_{s}(R) \) and \( f_{s}(R) \) are the accurate and the approximate
functions, respectively. One can see from Fig.\ 2 that the approximate
curves are very close to the exact ones at small \( R \). However, even at
quite large \( R \) the accuracy is not lower than 10\%. The total
accuracy is increasing with the nuclear charge \( Z \).

The triplet states of the two-electron atomic systems are always excited states.
Only the electron-nucleus coalescence line is formed for these states according
to the Pauli exclusion principle. It is important that the corresponding
wave function at the coalescence line \( F_{t}(R) \) behaves according 
to Eq.\ (29) like \( R^{2} \) as \( R \) approaches zero. Therefore, the function (46)
is not suitable in this case. Instead, for the lowest energy triplet state we
can propose the simple approximate function of the form
\begin{equation}\label{53}
f_{t}(R)=C_{t}R[\exp (-\lambda _{t}R)-\exp (-\beta _{t}R)],
\end{equation}
where \( \beta _{t}=\sqrt{-2E_{t}-Z^{2}} \) has to describe the behaviour of
\( F(R) \) at very large \( R \). It is easy to derive the following property
of this function at the origin
\begin{equation}\label{54}
\left\{ \frac{d}{dR}\left[ \frac{f_{t}(R)}{R^{2}}\right] \times
\left[ \frac{f_{t}(R)}{R^{2}}\right] ^{-1}\right\} \mid _{_{R=0}}
= - \frac{1}{2}\left ( \lambda _{t}+\beta _{t} \right )
\end{equation}
On the other hand, according to Eq.\ (29) the exact representation of \( F_{t}(R) \)
yields the value of \( (1/4-2Z/3) \) for this quantity. So
we obtain for the unknown exponent
\begin{equation}\label{55}
\lambda _{t} = \frac{4}{3} Z - \frac{1}{2} -\beta _{t}
\end{equation}
It is easy to verify that the inequality \( \lambda _{t}>\beta _{t} \) is
valid for all \( Z \). This condition is necessary to make the behaviour
of \( f_{t}(R) \) close to the correct one. Fig.\ 3 demonstrates
quite satisfactory behaviour of the approximate functions (53), especially for
\( R<5 \) a.u. We can observe that at small \( R \) the accuracy of
\( f_{t}(R) \) decreases with increasing \( Z \), whereas at large \( R \)
it increases with \( Z \).
The dips on the graphs of Fig.\ 2 and 3 are artifacts of the logarithmic scale,
since the logarithm of the absolute value of the difference of the two
functions goes to \( -\infty \) at the points of crossing the functions.
The overall accuracy therefore can be inferred only at the values of \( R \)
not too close to the dip.

\begin{figure}
\begin{center}
\epsfig{file=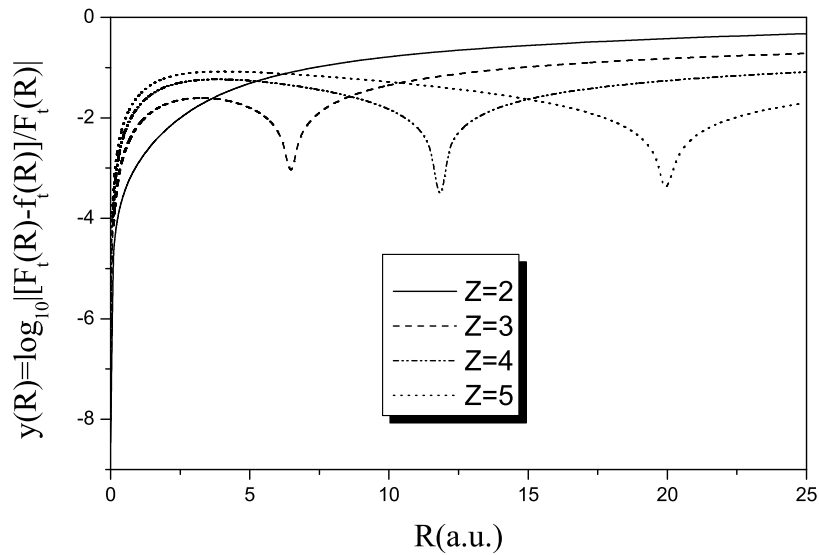,width=120mm}
\end{center}
\caption{Deviation of the approximate function \( f_{t}(R) \) from
the exact value \( F_{t}(R) \) for the \( 2S \)-triplet states of
the two-electron atom/ions with the nucleus charge \( Z=2,3,4,5 \).}
\label{fig3}
\end{figure}

The case of the electron-electron coalescence is the most complicated, 
because of the presence of the logarithmic term in the Eq.\ (43). However,
the general two-exponential form (46) can be applied in this case too,
but without the representation (47). 

For simplicity, let us put \( C=1/(1+\gamma ) \)
in Eq.\ (46). Then we obtain the approximate function \( f_{e}(R) \), which obeys
the condition \( f_{e}(0)=1 \), that in turn enables one to compare \( f_{e}(R) \)
with the expansion (43) in a simple manner. The exponent \( \beta  \)
representing the behaviour at very large \( R \) is equal to \( \beta _{e}=2\sqrt{-E} \)
and has to obey the inequality \( \lambda _{e}>\beta _{e} \), according to
the asymptotic representation (39). So, we have two unknown parameters \( \lambda _{e} \)
and \( \gamma _{e} \). The factor \( \gamma _{e} \) has to be positive,
because the ground state wave function is nodeless. Then, for the first derivative
\( f^{\prime }_{e}(R) \) at the origin we have
\( f^{\prime }_{e}(0)=-(\lambda _{e}+\beta _{e}\lambda _{e})/(1+\gamma _{e}) \).
Comparing this formula with Eq.\ (43), we obtain the first constraint:
\begin{equation}\label{56}
\lambda _{e} = 2Z + \gamma _{e}(2Z-\beta _{e}).
\end{equation}
It is easy to verify that the inequality \( 2Z>\beta _{e} \) is valid for all
\( Z\geq 1 \). Therefore, for the parameters \( \lambda _{e} \) and \( \gamma _{e} \)
obeying the constraint (56), the condition \( \lambda _{e}>\beta _{e} \) will
be valid for any positive \( \gamma _{e} \). To obtain the second constraint
for the parameters \( \lambda _{e} \) and \( \gamma _{e} \), one can use for example
an integral property of the exact wave function \( \Phi (R) \)
such as the normalization integral \( S=\int ^{\infty }_{0}\Phi ^{2}(R)R^{2}dR \)
with \( \Phi (0)=1 \). It is easy to calculate \( S \) using
the accurate CFHHM wave functions available. Replacing \( \Phi (R) \)
by \( f_{e}(R) \) in the integrand, and executing a simple integration, we
obtain the second constraint in the form
\begin{equation}\label{57}
\frac{1}{(2\lambda _{e})^{3}}+\frac{2\gamma _{e}}{(\lambda _{e}+\beta _{e})^{3}}
+\frac{\gamma _{e}^{2}}{(2\beta _{e})^{3}} = \frac{S}{2}(1+\gamma _{e})^{2}.
\end{equation}
The simplest way of solving the set of the equations (56), (57) is using the
well-known program Mathematica \cite{15}.  The equations
have a number of roots (including complex ones). However, only one root
turned out to be real and positive, and therefore it could be
applicable to \( \gamma _{e} \). The corresponding
solutions along with the other accompanying quantities, \( E, \) \( S \) and
\( \beta _{e} \) are presented in Table III.
It is seen from Table III that the approximate function parameters
fulfill all the conditions mentioned above.

\begin{table}[hbt]
\begin{center}
\caption{Parameters \( \beta _{e} \), \( \lambda _{e} \), and \( \gamma _{e} \)
of the approximate wave functions at the electron-electron coalescence line.
The accompanying values of the total energy \( E \) and the integrals \( S \)
are presented as well.}
\begin{tabular}{|c|c|c|c|c|c|}
\hline
\( Z \)&
\( -E \)&
\( S \)&
\( \beta _{e} \)&
\( \lambda _{e} \)&
\( \gamma _{e} \)\\
\hline
\hline
2&
2.9037244&
0.00452855&
3.40806&
5.54012&
2.60184\\
\hline
3&
7.2799134&
0.00127463&
5.39626&
8.32976&
3.8589\\
\hline
4&
13.655566&
0.000524535&
7.39069&
11.1232&
5.12583\\
\hline
5&
22.030917&
0.000264651&
9.38742&
13.9178&
6.39552\\
\hline
\end{tabular}
\end{center}
\end{table}

\begin{figure}
\begin{center}
\epsfig{file=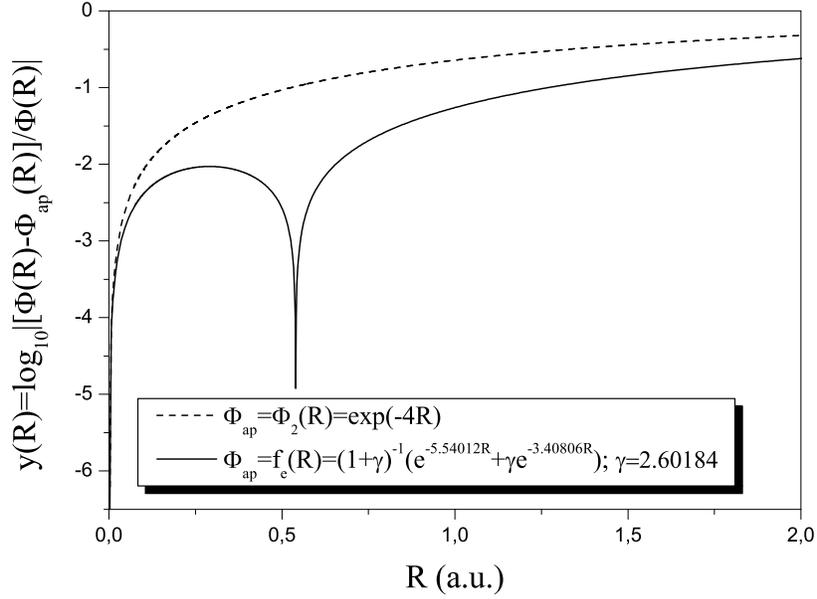,width=120mm}
\end{center}
\caption{Deviation of the approximate functions \( f_{e}(R) \) and
\( \Phi _{2}(R) \) from the exact wave function \( \Phi (R) \) at
the electron-electron coalescence line for the Helium atom.}
\label{fig4}
\end{figure}

The degree of coincidence for the accurate and approximate Helium wave functions
at the electron-electron coalescence line is presented in Fig.\ 4.
The solid line describes the deviation curve for the two-exponential approximate
function \( f_{e}(R) \) of the Helium atom. For comparison, we have also drawn
the corresponding one-exponential function (42), presented by the dotted line.
The graphs are limited by the value of \( R=2 \) a.u., because of the very
fast decay of the two-electron atomic wave functions at the electron-electron
coalescence line. For example, if \( \Phi (0)=1 \), then \( \Phi (2)\simeq 0.0006 \)
(for \( Z=2 \)). For comparison, we can point out that the corresponding value
of the wave function at the electron-nucleus coalescence line \( F_{s}(2)\approx 100\,\Phi (2) \).
The graphs on Fig.\ 4 demonstrate that the two-exponential function
is considerably more accurate than the one-exponential one, \( \Phi _{2}(R) \). All
of the approximate functions presented in this Section could be employed for the
evaluation of different atomic phenomena, and processes of the atomic photoionization
in particular.

\section{Conclusion}

We have considered and analyzed some particular solutions of the Schr\"{o}dinger
equation for the two-electron atom or ion (with the nucleus charge \( Z \)
and the total energy \( E \)) using the accurate CFHHM wave functions. We
have obtained mathematical relations between the partial derivatives taken at
the two-particle coalescence lines and the ordinary derivatives of the wave
function taken at the same coalescence lines. The relations were found for the
limit cases of very large and small distances \( R \) between one of the electrons
and the other electron close to the nucleus (electron-nucleus coalescence) or between
the two electrons close together and the nucleus (electron-electron coalescence). We
have obtained the only relation valid for all \( R\geq 0 \) and
\( Z\geq 1 \). It connects the first partial derivative on \( r_{1} \) (or
\( r_{2} \)) and the first ordinary derivative of wave function at the electron-electron
coalescence line.

We have examined both singlet and triplet \( S \)-states, and obtained the
asymptotic solutions (for very large \( R \)) with the exponents {\large }\( \left( -R\sqrt{-2E-Z^{2}}\right)  \)
or \( \left( -2R\sqrt{-E}\right)  \) {\large } for the electron-nucleus or electron-electron
coalescence, respectively. These results turned out to be valid for the ground
and singly excited bound states. We have derived the second order expansions
in \( R \) and \( \ln R \) in the vicinity of the triple coalescence point
for small \( R \). We have found that the Fock's logarithmic terms vanished
at the electron-nucleus coalescence line, unlike the case of the electron-electron
coalescence, but only in the framework of the second order expansion.

We have proposed simple two-exponential approximations for the two-electron
atomic wave functions at the coalescence points. The approximations are valid
for the lowest energy (nodeless) states. It was demonstrated that all of the
approximate functions
have quite satisfactory accuracy and could be employed at least for the initial
qualitative evaluation of a number of phenomena in the atomic physics.

\begin{acknowledgments}
We wish to thank Dr.\ E.~G.\ Drukarev for numerous discussions of the paper.
MYA is grateful to the Israeli Science Foundation (grant 174/03)
and Binational Science Foundation (grant 2002064). The researches of VBM and EZL were
supported by the Israeli Science Foundation, grant 131/00.
\end{acknowledgments}

%\newpage

\end{document}